# Quantum implementation of elementary arithmetic operations

G. Florio[1] and D. Picca[2]

INFN (Sezione di Bari) and Dipartimento di Fisica, Università di Bari,

Via Orabona 4, 70126, Bari, Italy

Abstract

Quantum computation has received great attention in recent years for its possible application to difficult problem in classical calculation. Despite the experimental problems of implementing quantum devices, theoretical physicists have tried to conceive some implementations for quantum algorithms. We present here some explicit schemes for executing elementary arithmetic operations.

**Introduction**

Quantum computation was considered one of the least appealing research field in physics. This stance has been drastically modified with the discovery of some powerful quantum algorithms (factoring of large numbers by P. Shor [1] and database search by L. Grover [2]). As a consequence, experimental and theoretical physics communities have shown a mounting interest in these topics. Great effort has been dedicated to the optimization of fundamental algorithms. Factoring has achieved important applications in cryptography (for military and security applications).

It is important to observe that the difficulty of manipulating large quantities of qubits (storing in memory registers, entanglement with the environment and decoherence) has limited the experimental development of a "quantum CPU"; this difficulty has also influenced the theoretical research: apart from some pioneer studies [3], the planning of a quantum processor is still at an early stage.

Notwithstanding the problems mentioned, it is interesting to explore, from a theoretical point of view, some possible circuit implementations of elementary arithmetic operations. This paper presents how some fundamental circuits can be implemented.

**1 - Quantum Adder**

The first circuit capable of executing sum was proposed by Barenco et al. [4]; it makes use of some supplementary memory registers to store information about the carry.

As observed by Draper [5], the use of the Quantum Fourier Transform (qFT) allows the calculus to be executed saving memory. In more detail, let us consider two integer numbers **a** and **b** stored in distinct registers, each constituted by n qubits; if the circuit for qFT is ap-plied to the ket $|a\rangle = |a_1 a_2 ... a_n\rangle$, the representative state of number **a**, its most significant qubit will be in a superposition of the form

$$|\varphi_1(a)\rangle = \frac{(|0\rangle + e^{2\pi i 0.a_1 a_2 ... a_n}|1\rangle)}{2^{1/2}} \qquad (1.1)$$

where the expression $0.a_1a_2..a_n$ represents a binary fraction; the other qubits will be in similar superposition with different binary fraction.

[1] E-mail address: giuseppe.florio@ba.infn.it

[2] E-mail address: picca@ba.infn.it

At this point, we apply the circuit shown in fig. 1.1 where we only use quantum phase gate $R_k$ with the form

$$R_k = \begin{pmatrix} 1 & 0 \\ 0 & e^{2\pi i/2^k} \end{pmatrix} \qquad (1.2)$$

For explaining clearly this procedure, we trace the state of the most significant qubit during the calculation:

$$|\varphi_1(a)\rangle \rightarrow \frac{1}{2^{1/2}}(|0\rangle + e^{2\pi i 0.a_1 a_2 ... a_n + 0.b_1}|1\rangle) \qquad \text{after } R_1 \text{ from } b_1$$

$$\rightarrow \frac{1}{2^{1/2}}(|0\rangle + e^{2\pi i 0.a_1 a_2 ... a_n + 0.b_1 b_2}|1\rangle) \qquad \text{after } R_2 \text{ from } b_2$$

..........................................

$$\rightarrow \frac{1}{2^{1/2}}(|0\rangle + e^{2\pi i 0.a_1 a_2 ... a_n + 0.b_1 b_2 ... b_n}|1\rangle) \qquad \text{after } R_n \text{ from } b_n$$

$$= |\varphi_1(a+b)\rangle$$

After applying this procedure to all qubits, the circuit for inverse qFT can be used to obtain the representative state of the number **a+b**. We note that there is no need for supplementary register to store the carry qubits.

The circuit of fig. 1.1 is very similar to the qFT; the number of gates needed to accomplish the operation can be estimated [5] in **$n\log_2 n$** (where n is the amount of qubits used); the main difference between qFT and quantum addition is the absence of Hadamard gates in the latter. As a consequence, all the operations in the circuit commute and thus they can be applied with an arbitrary sequence. In fig. 1.2, the scheme of the modified circuit is shown. We immediately note that all $R_1$ can be performed at the same time because they involve distinct qubits. This applies to $R_2$ and so on; as a result, the number of operations is the same but the time needed for the calculus is reduced by **n** times and so it is $O(\log_2 n)$.

**2 - Quantum Multiplier**

In this section, a circuit for multiplication will be explicitly derived.

This arithmetic operation can be considered as a kind of "extended sum". Let us consider two integer numbers **a** and **b**; the former is called "multiplicand", the latter is the "multiplier"; the multiplication algorithm basically consists in repeated sum of **a** for **b** times; the flowchart in fig. 2.1 shows this procedure explicitly.

The value of the multiplier is stored in a register labeled D, while register A is initially empty (i.e. it is in a fundamental state, for instance 0); first, the multiplicand is memorized in A and D is decremented by 1; if a fundamental value (0) is found in D, the calculus can be stopped; otherwise, the multiplicand is again added to A. This procedure is repeated until register D is empty (i.e. in the fundamental state).

The same scheme can also be used in the quantum domain.

The circuit in fig. 2.2 represents the general procedure for executing a multiplication. Let us consider two numbers **x** and **y** stored in the appropriate registers; let us suppose that both can be coded in n qubits; These numbers can now be considered as vectors in the respective computational bases and labeled by $\left|x\right\rangle$ and $\left|y\right\rangle$. The register marked with "Control" is composed by one qubit and is required for stopping the procedure. The register "Accumulator" must be composed by a number of qubits sufficient for storing the product: indeed, it is here that the operation results are progressively obtained. The controlled port D is applied when the control register is in the state $\left|0\right\rangle$ and decreases by 1 the value of the number coded in the state $\left|y\right\rangle$; its scheme is shown in fig. 2.3. Its action, when enabled, is analogous to that of the circuit for quantum sum; qFT and qFT$^{-1}$ represent respectively direct and inverse quantum Fourier transform; it is important to remark that in this case we use gate ($-R_k$) in which the phase is negative; an explicit matrix representation is

$$-R_k = \begin{pmatrix} 1 & 0 \\ 0 & e^{-\frac{2\pi i}{2^k}} \end{pmatrix} \tag{2.1}$$

An example will clarify this procedure. Let us consider the integer number 3: in a binary form it is expressed by 11 so it can be coded in state $\left|11\right\rangle$; we want to subtract integer 1 (coded in $\left|01\right\rangle$):

$$\left|11\right\rangle \xrightarrow{\text{qFT}} \begin{cases} \frac{1}{\sqrt{2}}(\left|0\right\rangle + e^{2\pi i(\frac{1}{2}+\frac{1}{4})}\left|1\right\rangle) \\ \frac{1}{\sqrt{2}}(\left|0\right\rangle + e^{2\pi i(\frac{1}{2})}\left|1\right\rangle) \end{cases} \xrightarrow{R} \begin{cases} \frac{1}{\sqrt{2}}(\left|0\right\rangle + e^{2\pi i(\frac{1}{2}+\frac{1}{4}-\frac{1}{4})}\left|1\right\rangle) \\ \frac{1}{\sqrt{2}}(\left|0\right\rangle + e^{2\pi i(\frac{1}{2}+\frac{1}{2})}\left|1\right\rangle) \end{cases} =$$

$$= \begin{cases} \frac{1}{\sqrt{2}}(|0\rangle + e^{2\pi i(1/2)}|1\rangle) \\ \frac{1}{\sqrt{2}}(|0\rangle + e^{2\pi i(1)}|1\rangle) \end{cases} \xrightarrow{\text{qFT}^{-1}} |10\rangle$$

We have obtained state $|10\rangle$ that represents integer 2. this procedure can be repeated until we obtain state $|00\rangle$ obviously representing 0. Thus we can set to zero the multiplier register.

Another circuit block is the doubly controlled gate "S". Its action is that of summing the multiplicand to the contents of the accumulator by using the quantum adder.

C-NOT gates after D gates control if the contents in the $|y\rangle$ register are $|00....0\rangle$; in this case, the control qubit passes from state $|0\rangle$ to state $|1\rangle$ disabling all the next D and S gates and stopping the multiplication; actually the D and S gates are not two-qubits gates; their activation is controlled by all the qubits in the $|y\rangle$ register. The complete scheme is in fig. 2.3

Let us trace the procedure step by step.

Accumulator and control registers are loaded respectively with states $|00....0\rangle$ (the number of qubit depends on the resources needed to store the product) and $|0\rangle$; $|x\rangle$ and $|y\rangle$ registers are used to store the multiplicand and multiplier. qFT is applied to the accumulator to prepare the system to repeated sums; the first C-NOT gate is needed to check $|y\rangle$ and verify if it is in the fundamental state $|00....0\rangle$; if not, Controlled-S gate is enabled and $|x\rangle$ is stored in the accumulator in the Fourier form; precisely we have

$$\frac{(|0\rangle + e^{2\pi i 0.x_1 x_2 \dots x_n}|1\rangle)......(|0\rangle + e^{2\pi i 0.x_{n-1}x_n}|1\rangle)(|0\rangle + e^{2\pi i 0.x_n}|1\rangle)}{2^{n/2}} \tag{2.2}$$

The state $|y\rangle$ is transformed into $|y-1\rangle$ by the D gate and another C-NOT check the state in the register; if it is still different from the fundamental, the procedure is repeated summing again $|x\rangle$ to the accumulator; we obtain (for simplicity we show only the most significant qubit):

$$\frac{1}{\sqrt{2}}(|0\rangle + e^{2\pi i(x_1/2 + x_2/4 + \dots + x_n/2^n + x_1/2 + x_2/4 + \dots + x_n/2^n)}|1\rangle) \tag{2.3}$$

which represents the most significant qubit of the state $|2x\rangle$.

The procedure will continue until the $|y\rangle$ register reaches zero; C-NOT is then enabled and therefore the control qubit is commuted from $|0\rangle$ to $|1\rangle$; next S and D gates are disenabled; last gate (qFT$^{-1}$) is applied to the accumulator register extracting the state $|x \bullet y\rangle$. As a result, the product is obtained.

**Conclusions**

The analysis has shown how to build circuits for quantum sum and multiplication. It must be observed that, in both cases, inputs are vectors of the computational bases and so for outputs; an important consequence is that, notwithstanding the fact that our procedures make use of superpositions of states and sum of phases (typically in the qFT), the final states do not have the typical ambiguity of quantum mechanics. The measure of an observable allows to obtain one of the eigenvectors of the hermitian operator associated to it; in our case, the output is already a vector of the computational basis and hence the measured result is deterministic.

The circuits implemented can be used to build a more complex scheme which can execute more complicated operations. As will be shown in an upcoming paper, the circuits in question are the basis of the ALU of a primitive quantum CPU.

Some comments about our proposal are in order. We have completely neglected decoherence and control problems during the evolution of the system. This choice was suggested by the present state of research in quantum computation. Experimental works show how difficult is to manage a few real qubits. As a consequence, our circuits are hard to obtain by means of present technology. Our devices should be considered "sneak peeks" to future computers.

**References**

1. P.W. Shor, *Algorithms for quantum computation: discrete logarithms and factoring; in Proceedings of the 35$^{th}$ Annual Symposium on the Foundations of Computer Science*, IEEE Computer Society Press, Los Alamitos, CA (1994).
2. L. Grover, *Proceedings of 28$^{th}$ Annual ACM Symposium on the Theory of Computation*, 212-219, ACM Press, N.Y. (1996). L. Grover, *Quantum mechanics helps in searching for a needle in a haystack*, Phys. Rev. Lett., 79(2): 325, www.arxiv.org quant-ph/9706033 (1997).
3. M.A. Nielsen, I. Chuang, *Programmable quantum gate arrays*, www.arxiv.org quant-ph/9703032 (1997). A.M. Wang, *An universal quantum computer - quantum CPU*, www.arxiv.org quant-ph/9910089 (2000). M. Hillery, V. Buzek, M. Ziman, *Probabilistic implementation of universal quantum processors*, www.arxiv.org quant-ph/0106088 (2001). M. Hillery, V. Buzek, M. Ziman, *Implementation of quantum* *maps by programmable quantum processors*, www.arxiv.org quant-ph/0205050 (2002). Y. Yu, J. Feng, M. Zhan, *Multi-output programmable quantum processor*, www.arxiv.org quant-ph/0209069 (2002). A. Vlasov, *Quantum processors and controllers*, www.arxiv.org quant-ph/0301147 (2003).

4. V. Vedral, A. Barenco, A. Ekert, *Quantum networks for elementary arithmetic operations*, www.arxiv.org quant-ph/9511018 v1 (1995).
5. T.G. Draper, *Addition on a quantum computer*, www.arxiv.org quant-ph/0008033 (2000).

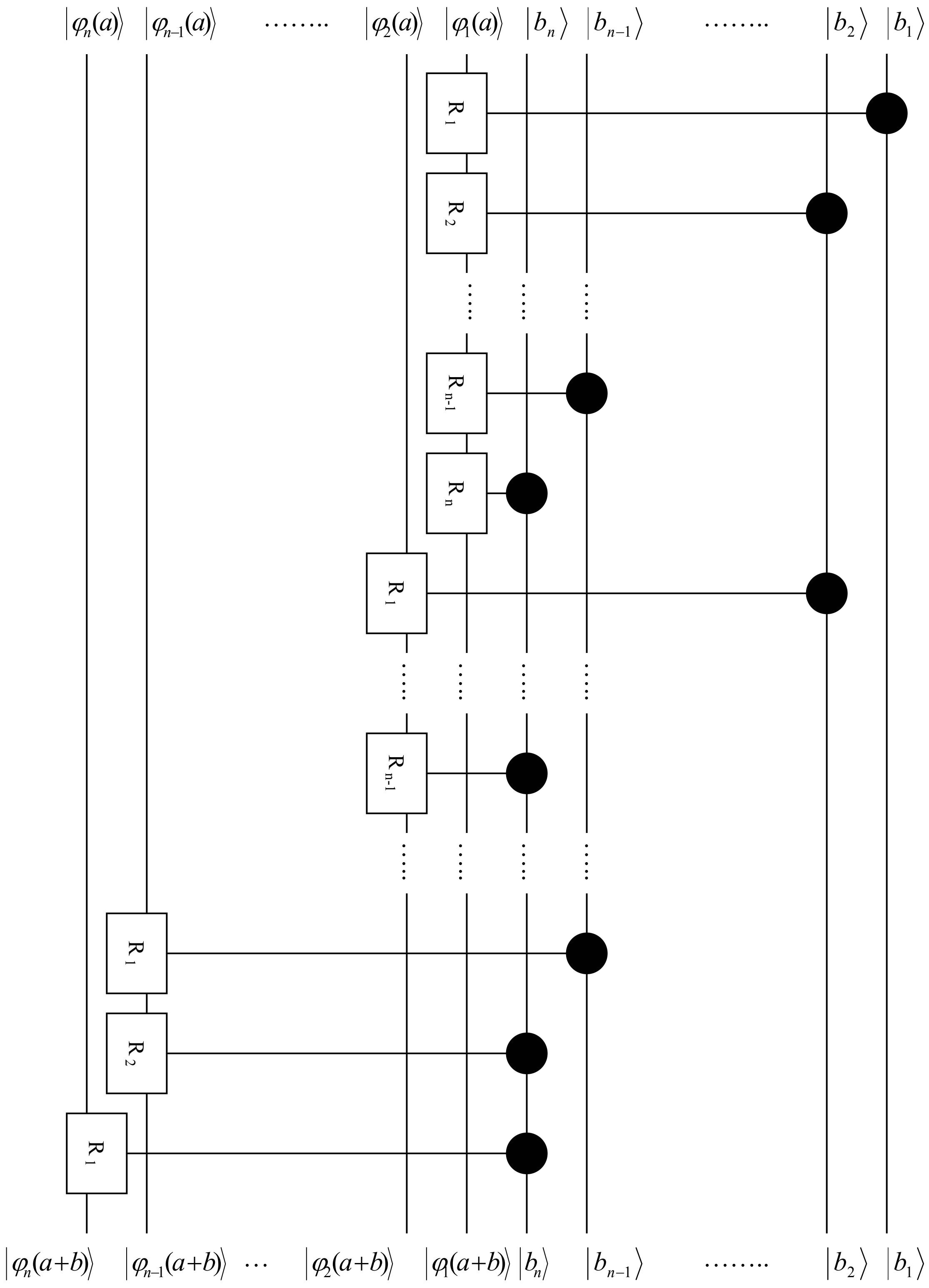

**Fig. 1.1** – Quantum Adder

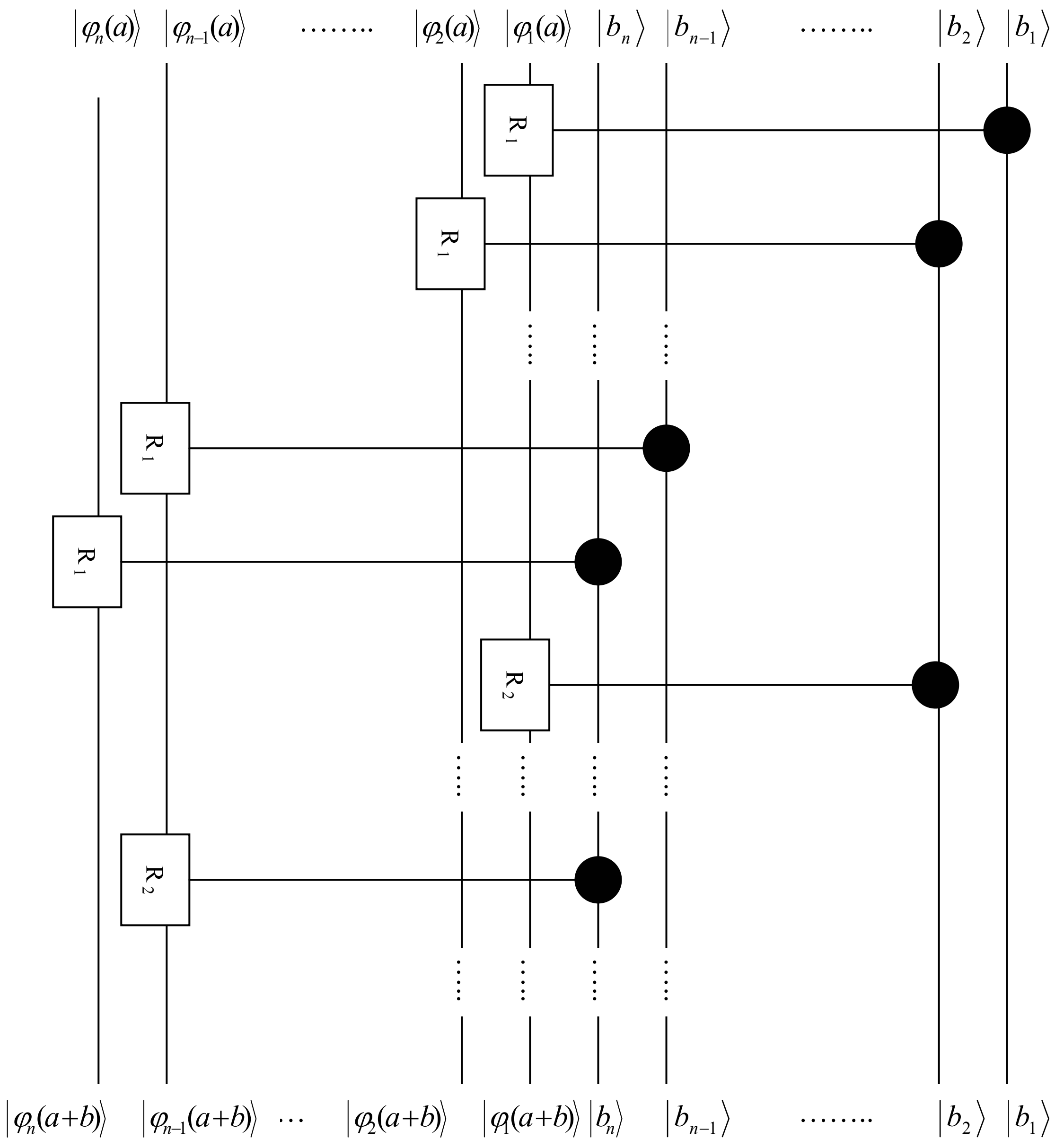

**Fig. 1.2** – Parallel Quantum Adder

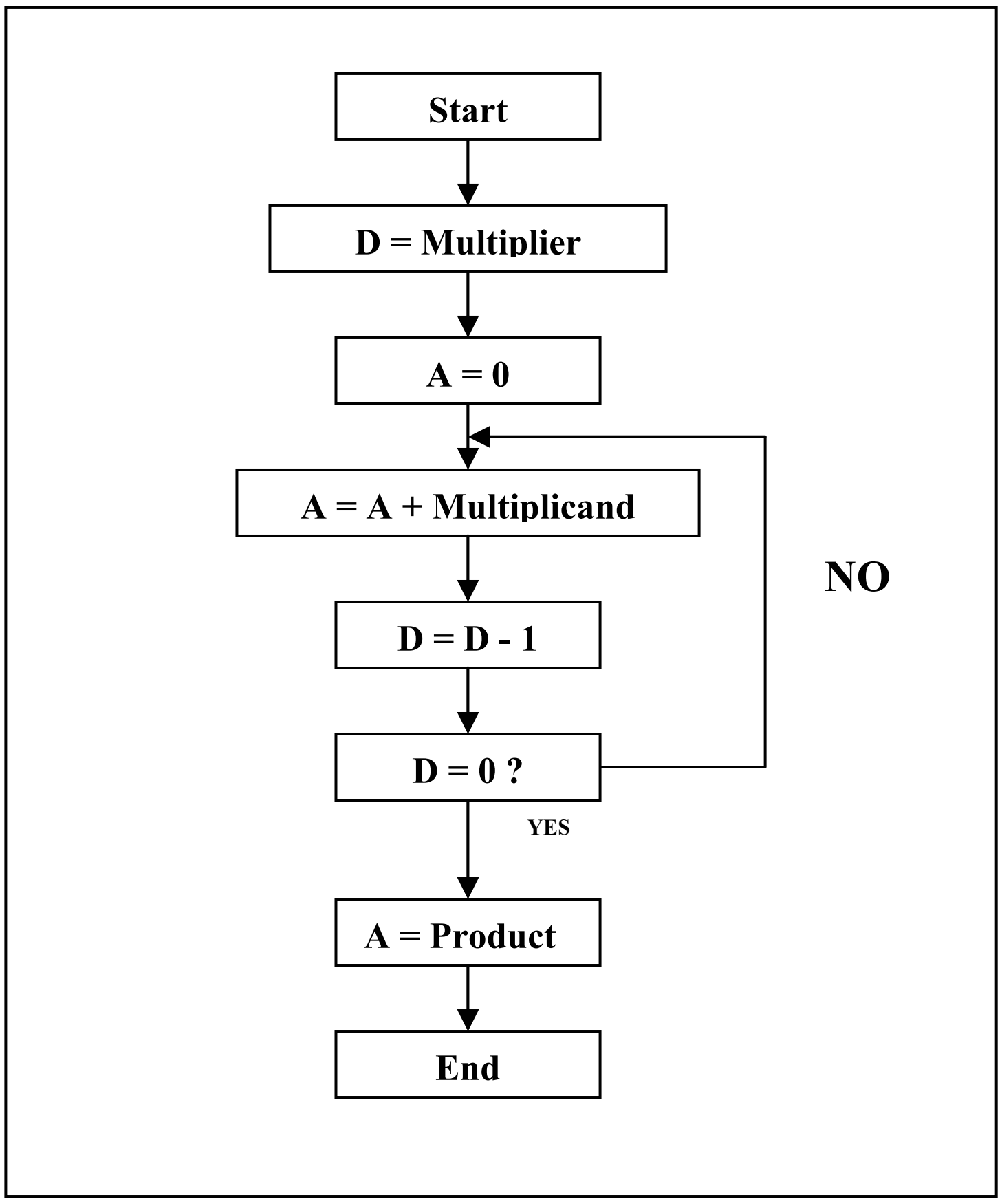

**Fig. 2.1** – Multiplication Flowchart

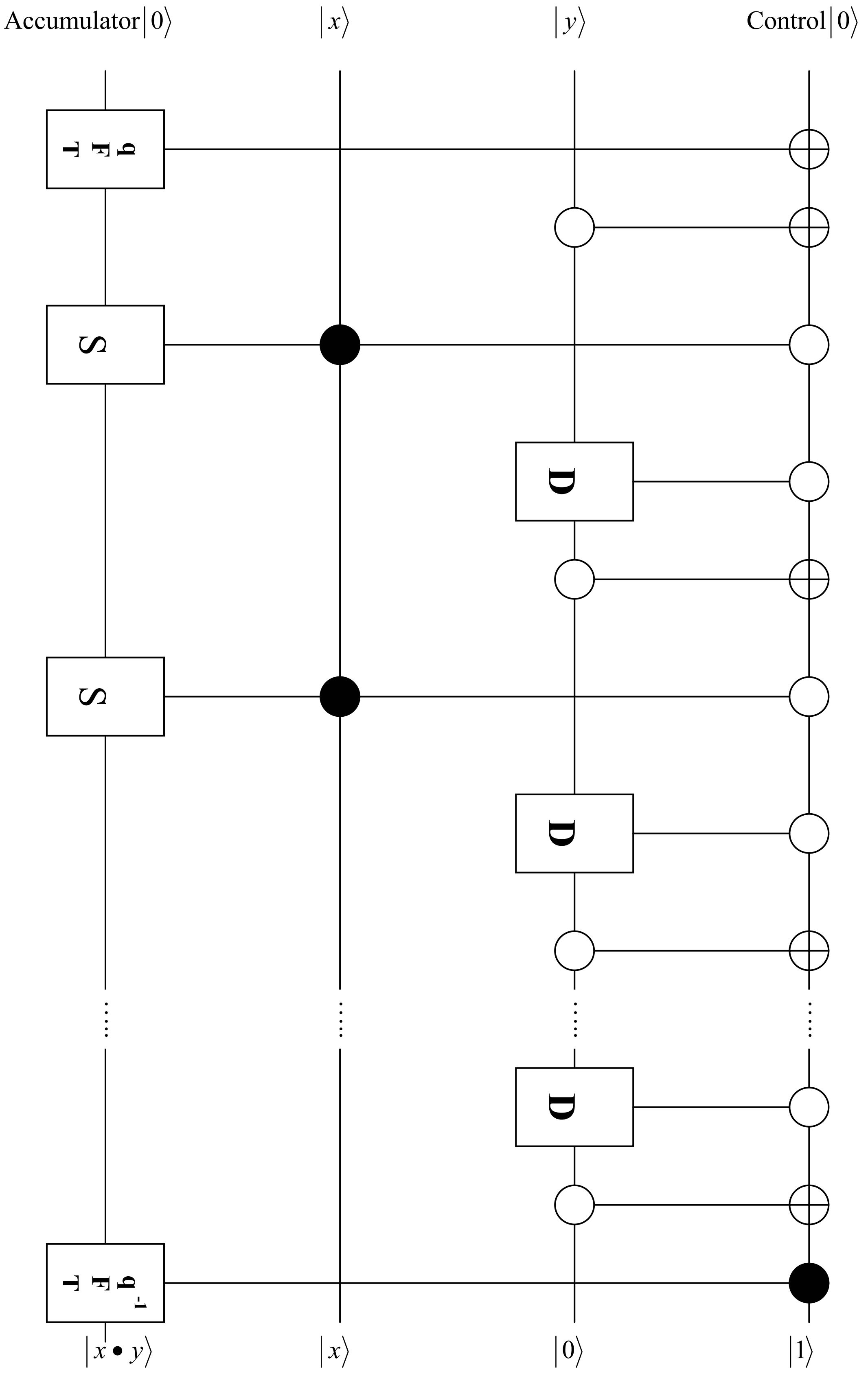

**Fig. 2.2** – Quantum Multiplier

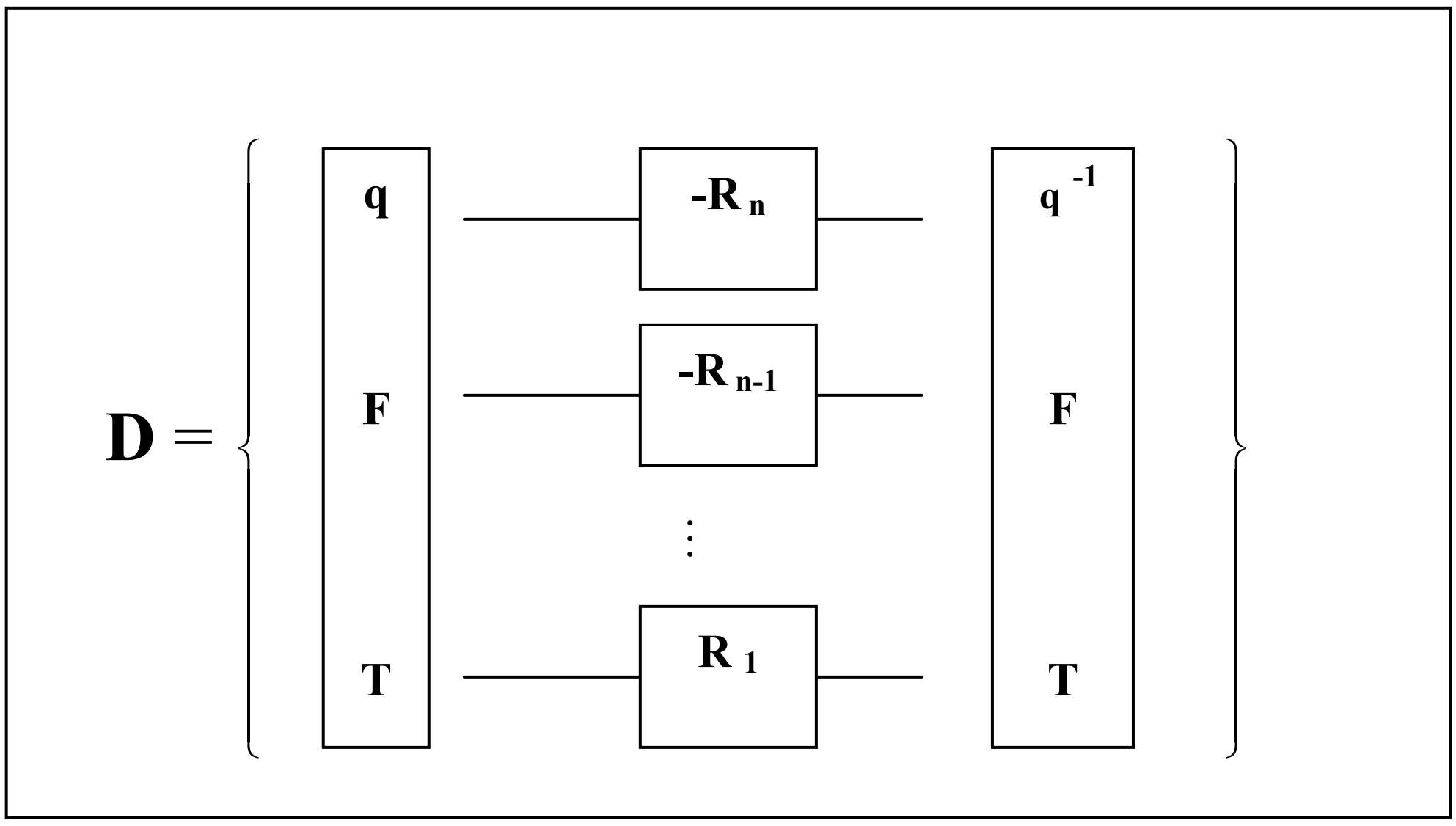

**Fig. 2.3** – Decrement Gate